\documentclass[useAMS,usenatbib]{mn2e}

\usepackage{amsmath}
\usepackage{amssymb}
\usepackage{graphicx}
\usepackage{subfigure}

\begin{document}

\title{The host galaxies of radio-loud AGN: colour structure}

\author[E. J. A. Mannering et al.]
{E. J. A.~Mannering$^{1}$\thanks{E-mail: l.mannering@bristol.ac.uk}, D. M.~Worrall$^{1}$, M.~Birkinshaw$^{1}$ \\
$^{1}$H.H. Wills Physics Laboratory, University of Bristol, Tyndall Avenue,  Bristol BS8 1TL, UK } 

\maketitle

\label{firstpage}

\begin{abstract}
We construct a sample of 3,516 radio-loud host galaxies of active galactic nuclei (AGN) from the optical Sloan Digital Sky Survey (SDSS) and Faint Images of the Radio Sky at Twenty cm (FIRST). These have 1.4\,GHz luminosities in the range $10^{23}-10^{25}$\,WHz$^{-1}$, span redshifts $0.02<z<0.18$, are brighter than $r^{*}_{petro}<17.77$\,mag and are constrained to `early-type' morphology in colour space  ($u^{*}-r^{*}>2.22$\,mag). Optical emission line ratios (at $>3\sigma$) are used to remove type 1 AGN and  star-forming galaxies from the radio sample using BPT diagnostics. For comparison, we select a sample of 35,160 radio-quiet galaxies with the same $r^{*}$-band magnitude-redshift distribution as the radio sample. We also create comparison radio and control samples derived by adding the NRAO VLA Sky Survey (NVSS) to quantify the effect of completeness on our results. 

We investigate the effective radii of the surface brightness profiles in the SDSS $r$ and $u$ bands in order to quantify any excess of blue colour in the inner region of radio galaxies.   We define a ratio $R=r_{e}(r)/r_{e}(u)$ and use maximum likelihood analysis to compare the average value of $R$ and its intrinsic dispersion between both samples. $R$ is larger for the radio-loud AGN sample as compared to its control counterpart, and we conclude that the two samples are not drawn from the same population at $>99\%$ significance. Given that star formation proceeds over a longer time than radio activity, the difference suggests that a subset of galaxies has the predisposition to become radio loud. 
We discuss host galaxy features that cause the presence of a radio-loud AGN to increase the scale size of a galaxy in red relative to blue light, including excess central blue emission, point-like blue emission from the AGN itself, and/or diffuse red emission. We favour an explanation that arises from the stellar rather than the AGN light.

\end{abstract}

\section{Introduction}
Considerable uncertainties remain as to
what controls the apparent link between the activity of active galactic nuclei (AGN) during the time that a black hole is fed and star formation.  The
accretion disk surrounding a super-massive black hole (SMBH) emits highly energetic radiation
and particles, and can form powerful winds and/or collimated,
relativistic jets \citep[e.g.,][]{Breugel}. The radiation and
outflows might then affect the interstellar medium, triggering star
formation which might be detectable as an excess of blue light from the central regions of galaxies. Alternatively, the activity of the AGN may be
sparked by specific events in the galaxy's past.  For example, there
is morphological evidence that activity in radio galaxies
might be triggered by mergers and galaxy interactions
\citep[e.g.,][]{Gonzalez}, which in turn could contribute to central
blue light through enhanced star formation.  In either case, we
might expect an observable association between AGN activity, for which
in this work we use radio loudness as a tracer, with bluer stellar continuum in the central regions of the galaxy.

Previous work has hinted at such a relationship.  For example,
\citet{Mahab} examined the central light of 30 radio galaxies
as compared to 30 normal galaxies from the Molongo
Reference Catalogue and found excess blue light in the inner regions of the radio galaxies as compared with the control sample.
This central excess is unrelated to the conventional colour gradient of the broader distribution of stars. \citet{Gonzalezperez} studied colour variations within galaxies from the Sloan Digital Sky Survey \citep[SDSS;][]{York} DR7, finding marginally steeper colour gradients in massive galaxies with nuclear activity and concluded that this is due to a higher fraction of young stars in their central regions.

In the present paper we expand the work of \citet{Mahab}, by combining optical information from SDSS with the FIRST and NVSS radio surveys \citep{Becker, Condon98} to construct a local ($0.02<z<0.18$) sample of powerful radio-loud AGN host galaxies (R-AGN), and a control sample of `normal' early-type ellipticals.
We define $R$ as the ratio of SDSS $r^{*}$ to $u^{*}$ band de Vaucouleurs effective radii and compare its value between the samples. 

The paper is structured as follows. In Section \ref{sec:Surveys} we briefly describe the surveys used to construct our samples. Separation into the radio-loud and control samples is discussed in Section \ref{sec:sampleselection}. Section \ref{sec:analysis} outlines the maximum likelihood analysis used to compare the average values of $R$ for the two samples. In Section \ref{sec:Discussion} we present a discussion of our findings in the context of radio-AGN fuelling mechanisms and the properties of the host galaxies. 
Throughout, we use a flat $\Lambda$CDM cosmology with $\Omega_{m_{0}}=0.3$ and $\Omega_{\Lambda 0}=0.7$. We adopt $H_{0}=70$ km\,s$^{-1}$\,Mpc$^{-1}$.

\section{Data Sources}
\label{sec:Surveys}
\subsection{SDSS}
SDSS is an imaging and spectroscopic survey covering $\sim$10,000 deg$^{2}$, primarily in the northern hemisphere. Its 7th data release \citep[SDSS-DR7;][]{DR7} contains over 350 million entries, of which about one million are confirmed galaxies with spectroscopic follow up.

In the photometric sample, flux densities are measured simultaneously in five broadband filters ($u,g,r,i,z$), with effective wavelengths $3551$, $4686$, $6165$, $7481$ and $8931$\AA\, \citep{Fukugita}. 
All magnitudes are given on the AB$_{v}$ system \citep{Oke}, and are determined using Petrosian apertures \citep[see][for a complete definition of the Petrosian system]{Blanton, Graham}. In common with \citet{Ivezic2002}, we refer to SDSS measured magnitudes as  $u^*,g^*,r^*,i^*$ and $z^*$ due to the uncertainty in the absolute calibration of the SDSS photometric system, which has been assessed at $\lesssim$0.03\,mag \citep{Stoughton}. 
 For a complete guide to the photometric system, see \citet{Fukugita}, \citet{photo} and \citet{Stoughton}.

\subsubsection{Main galaxy sample (MGS)} 
\label{subsubsec:mgs}
We use the main galaxy survey spectroscopic sample \citep[MGS, ][]{Spectarget}. 
The MGS algorithm selected extended sources brighter than $r^{*}_{petro} = 17.77$\,mag with Petrosian half-light surface brightness $\mu_{50} \leq 24.5$\,mag arcsec$^{-2}$, providing $\sim$ 90 galaxies deg$^{-2}$ \citep{Blanton}. Candidate galaxies were then observed spectroscopically. $99.9\%$  of the resulting redshifts have velocity errors $<$\,30\,km\,s$^{-1}$. The survey is unbiased, except for galaxies with close companions. Around $6\%$ of galaxies satisfying the photometric target criteria were not included for spectroscopic follow up due to a companion galaxy within the 55\arcsec\ minimum fibre separation, although some of these galaxies have been subsequently observed.
The MGS contains $\sim$680,000 spectroscopically confirmed galaxies. The full target selection is described in \citet{Spectarget}. 

We corrected for Galactic extinction using the SDSS `reddening' parameter, which was derived from maps of the infrared emission from dust across the sky in accordance with \citet*{SFD98}. 
We applied $k$-corrections using the SDSS-derived `kcorr' parameter as detailed in \citet{BlantonK}. Typical corrections $k_{u}, k_{g}, k_{r} $ were $(0 - 0.4)$, $(0 - 0.3)$, and $(0 - 0.1)$\,mag.

\subsection{FIRST}
\label{subsec:FIRSTSURVEY}
The FIRST survey \citep{Becker} utilized the VLA (at 1.4\,GHz in B array) to map the radio sky over $\sim$9000\,deg$^{2}$ in the Northern hemisphere and in a $2\overset{^{\circ}}{.}5$ wide strip along the celestial equator. It has $\sim 8400$\,deg$^{2}$ overlap with SDSS, contains $\sim 97$ sources deg$^{-2}$ at the 1\,mJy survey threshold and reaches an rms sensitivity of $\sim$0.15\,mJy\,beam$^{-1}$.
In this configuration, the VLA has a synthesized beam of 5.4\arcsec\ FWHM, providing accurate flux densities for small-scale radio structures, but underestimating the flux densities of sources extended to several arcminutes. At the 1\,mJy survey threshold, individual sources have 90\% confidence astrometric errors $\lesssim 1\arcsec$. We adopted a mean spectral index of $\alpha=0.7$ (where $S_{\nu} \propto \nu^{-\alpha}$) and obtained rest-frame 1.4\,GHz power densities by applying k-corrections assuming the usual form 

\begin{equation}
L_{\nu, rest}=(1+z)^{(\alpha-1)}S_{\nu}\,4\pi D_{L}^{2}
\end{equation} 

\noindent where $D_{L}$ is the luminosity distance. 
\subsection{NVSS} 

The NRAO VLA Sky Survey \citep[NVSS,][]{Condon98} mapped the radio sky (at 1.4\,GHz in D array) north of $-40\overset{^{\circ}}{ }$ declination. The survey is complete down to a point source flux density of $\sim$\,2.5\,mJy. The synthesized beam of 45\arcsec\ is much larger than FIRST, providing more accurate flux measurements for highly extended sources. Astrometric accuracy ranges from 1\arcsec\ for bright sources to 7\arcsec\ for the faintest detections. The entire survey contains over 1.8 million unique sources brighter than 2.5\,mJy. $k$-corrections are applied as in \S\ref{subsec:FIRSTSURVEY}.

\section{Sample selection}
\label{sec:sampleselection}

\subsection{Cross-correlation: FIRST-SDSS}
\label{subsec:Crosscorrelation}
\begin{figure}
\begin{center}
\flushleft
\includegraphics[width=\columnwidth]{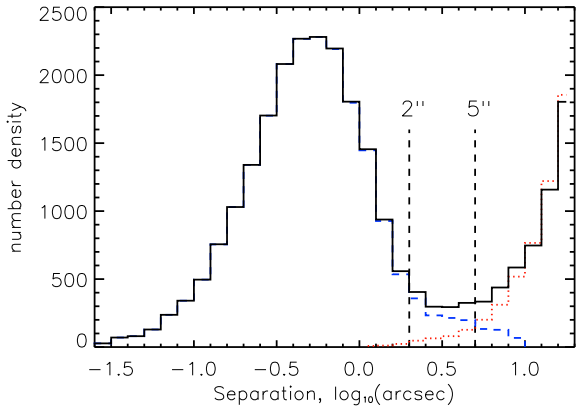}  
\caption{The distribution of angular separations of the matches between FIRST and SDSS, truncated at a 20\arcsec\ radius (black). The distribution of matches between the FIRST positions offset by $1^{\circ}$ and SDSS is shown as a red dotted line. This indicates the number of likely false matches. The net distribution of `real' matches is shown by the blue dashed line, and our 2\arcsec\ cut indicates 99.9\% efficiency of the matched sample.}
\label{fig:idlsep}
\end{center}
\end{figure}

We initially derived a FIRST-SDSS sample of radio galaxies and a comparison control sample with no nearby detectable radio counterpart. 
Figure \ref{fig:idlsep} shows the distribution of angular separations between SDSS objects and their nearest FIRST counterpart. We find that at $\sim$\,5\arcsec\ the distribution becomes dominated by random matches. Hence we adopt a match criterion of 2\arcsec\ angular separation to avoid such false matches.
Our cross-matched catalogue contains $25,931$ galaxies and we denote this as our preliminary `radio' sample. All unmatched sources are initial candidates for our control sample of radio-quiet galaxies. 

We define our efficiency as the fraction of matches in our radio sample which are physically real (including contamination from line of sight false matches, which cannot be accounted for here). In order to evaluate the number of random false matches in our sample, we offset the RA and DEC of the FIRST sources by $1^{\circ}$ and re-matched to our set of $680,056$ SDSS galaxies, selecting objects within a 2\arcsec\ radius.
We found on average $58$ random matches, corresponding to 0.3\% contamination of the FIRST-SDSS match sample ($>$99\% efficiency, Figure \ref{fig:idlsep}, red dotted line). 

We also need to know the completeness of our matching criteria - the fraction of correct matches we recover. A factor affecting this is the possible exclusion of large lobe-dominated radio sources \citep{Kimball}. Both lobes will be included in FIRST but may be excluded in the cross-matched subset for lobe-core distances $>$\,2\arcsec, if the core is weak. We estimated $\sim$\,8$\%$ of real matches between the SDSS subset and FIRST that are excluded from our radio sample, by assuming the radio lobes are outside a 2\arcsec\ radius from the optical core, but within a 5\arcsec\ radius (see Figure \ref{fig:idlsep}). 
Our matching algorithm efficiency and completeness are similar to \citet{Kimball}, who match FIRST to SDSS-DR6 positions within a 2\arcsec\ radius for sources with $z\leq$ 2 (95\% efficiency and 98\% completeness\footnote{\citet{Kimball} estimate their completeness and efficiency by fitting a gaussian to the nearest-neighbour distribution (representing physical matches) plus a rising linear function (representing random matchings).}).

However, we could not account for sources with FIRST lobes $>5$\arcsec\ from the optical core or the population of FIRST double-lobed sources with no detectable radio core. \citet{Ivezic2002} cross-correlated all FIRST sources with SDSS, then identified potential double-lobed radio sources with undetected cores\footnote{Via comparing the mid-points of FIRST pairs to SDSS sources within a separation $<\,90$\arcsec\, and accepted all matches with offsets $<\,3$\arcsec.}, estimating these contribute less than $10\%$ of all radio sources \citep[][estimate $\sim 5\%$ of their radio-loud AGN sample had no FIRST detection]{Best2005a}.

\begin{figure}
\centering
\includegraphics[width=0.45\textwidth]{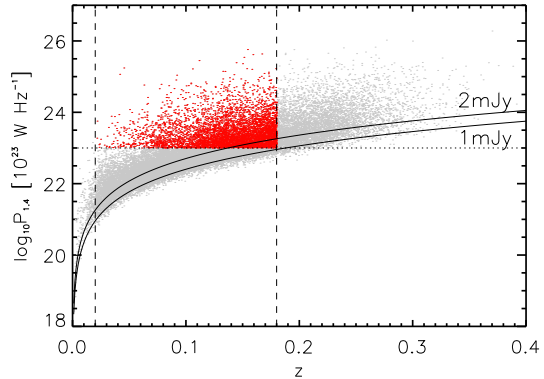}
\caption{Integrated FIRST radio power vs redshift for the preliminary `radio' sample (see text). The solid black lines trace peak radio power at the 1\,mJy detection threshold and at 2\,mJy, the dashed lines denote the redshift limits imposed in \S\ref{subsec:redshiftrange} and the dotted line denotes the radio power cut imposed in \S\ref{sec:radioagn}. The red points are the remaining radio sample after these cuts are applied.}
\label{fig:radiopower_z}
\end{figure}

\citet{Becker} estimated the FIRST catalogue to be $95\%$ complete at 2\,mJy and $80\%$ complete down to the survey limit of 1\,mJy. 
Figure \ref{fig:radiopower_z} shows integrated radio power against redshift for the preliminary radio sample (25,931 sources). The solid black lines trace the 1\,mJy and 2\,mJy peak flux density thresholds, the dotted and dashed lines show cuts we applied in redshift and radio power (see \S\ref{subsec:redshiftrange} and \S\ref{sec:radioagn}), and the red points are the resultant selection of radio sources. It is noted that for sources that are not well-described by an elliptical Gaussian model, the integrated flux density as derived by FIRST may be an inaccurate measure of the true value. Such sources with radio powers corresponding to flux densities below the 1\,mJy threshold can be seen in Figure \ref{fig:radiopower_z}.

We note that although the two cuts improve the sample completeness (all red points are above 1\,mJy), many of our sources are below the 2\,mJy line, and so our sample completeness cannot be $>80\%$. 
Combining this with our estimation of completeness in our selection criteria ($\sim 90\%$), we estimate the actual completeness of our preliminary radio sample to be $\sim 72\%$ for sources brighter than 1\,mJy. 
The incompleteness is dominated by statistical effects rather than the $10\%$ effect of sources missing due to weak cores.

To produce a preliminary control sample, we select sources from the SDSS subset which fall within the same spatial region as FIRST ($\sim8000$ deg$^{2}$). 
We selected all objects from this subset which are not matched to a FIRST source within 2\arcsec\ of their optical core, providing $\sim$625,664 radio-quiet galaxies. The efficiency of this sample, based on Figure \ref{fig:idlsep} is $>99.9\%$ (i.e., $<1\%$ contamination by radio sources detectable at the FIRST flux density limit) and the sample completeness is $>92\%$.

\subsection{Cross-correlation: FIRST-NVSS-SDSS}
\label{subsec:crossNVSSFIRST}

For FIRST sources substantially larger than $5$\arcsec\ some flux density is resolved out, leading to an increase in the survey threshold for extended objects and flux density underestimates for larger objects \citep{Lu}. 
Our FIRST-SDSS radio sample is large due to the number of FIRST candidate sources near the 1\,mJy survey limit. Its limitations are potential flux density underestimates, and incompleteness to extended radio sources, notably extended, double-lobed sources with no detectable radio core within several arcseconds of its optical counterpart.

To attempt to quantify these limitations, we created a complementary sample of radio-loud AGN host galaxies using both NVSS and FIRST. \citet{Best2005a} note that the 45\arcsec\ resolution of NVSS is large enough that $\sim 99\%$ of all radio sources are contained within a single component, allowing for a higher sample completeness. However, NVSS is less deep than FIRST and consequently the matched sample contains fewer galaxies. The NVSS-FIRST-SDSS matched sample therefore does not benefit from such good statistics.

We used the Unified Radio Catalogue constructed by \citet{Kimball} (see sample `C' in their table 8), selecting sources which are detected by both FIRST and NVSS, matched to within 25\arcsec. 
This selection yields a radio flux density limited ($S_{\mathrm{1.4GHz}}>2.5$\,mJy) catalogue (hereafter FIRST-NVSS) containing 141,881 sources. We cross-matched these to the $\sim680,056$ MGS sources adopting the same 2\arcsec\ match criterion used in \S\ref{subsec:Crosscorrelation}. Our cross-matched catalogue contains 5,719 galaxies and we denote this as the preliminary `comparison radio' (hereafter CR) sample. Integrated flux densities as derived by \citet{Kimball} are adopted in our analysis. 

We estimated our matching criterion of 2\arcsec\ to be $>99\%$ complete, and $>99\%$ efficient (8 random matches were found when the FIRST-NVSS RA and DEC were offset by $1^{\circ}$). The FIRST-NVSS catalogue used to create the CR sample is 99$\%$ complete and matched with $96\%$ efficiency \citep[see table 2 of ][]{Kimball}. We therefore estimated the comparison radio sample to be $>99\%$ complete and $>95\%$ efficient (see Table \ref{tab:comeff}). 5 of the 5,719 galaxies in the CR sample reside in the control sample derived in \S\ref{subsec:Crosscorrelation} and these were removed. 4,935 of the CR sample (86$\%$) are also in the FIRST-SDSS sample.

\begin{table}
\centering
{\small
\begin{tabular}{p{1.7cm} p{1.2cm}p{0.8cm}p{1.5cm}p{1.5cm}}  
\hline
\hline
Matched &$S_{\mathrm{lim}}$(mJy) &Objects&Completeness&Efficiency \\
surveys & & & &\\ [0.5ex]    
\hline   
FIRST-SDSS & 1.0 & 25,931 & $72\%$ & $>99\%$ \\
FIRST-NVSS-SDSS & 2.5 & 5,719 & $99\%$ & $>95\%$ \\
[0.5ex] 
\hline
\end{tabular} 
}

\caption{Radio flux density limits, $S_{\mathrm{lim}}$, of both the radio sample (FIRST-SDSS) and the candidate CR sample (FIRST-NVSS-SDSS), the number of objects in each and the estimated completeness and efficiency.} 
\label{tab:comeff} 
\end{table}

\subsection{Redshift range}
\label{subsec:redshiftrange}
In the following sections (\S\ref{subsec:redshiftrange}-\ref{subsec:redshiftdist}) we discuss the secondary selection criteria applied to the FIRST-SDSS sample. The CR sample follows the same path, and a summary of the final CR sample properties are presented alongside those of the final radio sample in \S\ref{subsec:crprop}.

\begin{figure}
\begin{center}
\flushleft
\includegraphics[width=\columnwidth]{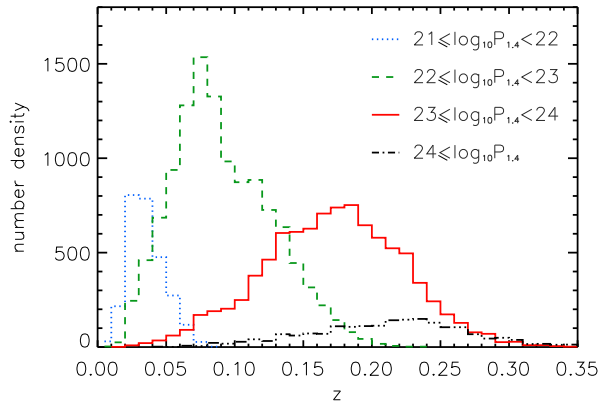}  
\caption{Distributions of FIRST-SDSS radio sample redshifts for $\mathrm{log}_{10}$P$_{\mathrm{1.4\,GHz}}$ luminosity bins. We cut at $0.02<z<0.18$, providing a high sample completeness ($>98\%$) down to P$_{\mathrm{1.4\,GHz}} \approx 10^{23}$\,W\,Hz$^{-1}$.}
\label{fig:lumredradio}
\end{center}
\end{figure}

We imposed a redshift cut of $0.02<z<$ 0.18 to ensure that the galaxy spatial structure could be well examined. For $z<0.02$, redshift is not a reliable distance indicator. Figure \ref{fig:lumredradio} shows less than 2\% of matched sources have redshift $<0.02$.
The MGS magnitude limit of $r^{*}_{petro}<$17.77 corresponds to L$_{r^{*}}> 2.6\times10^{23}$\,W\,Hz$^{-1}$ at $z=0.18$, and our sample should be complete to this optical luminosity density. 

We restricted the SDSS-derived redshift confidence ({\fontfamily{pcr}\selectfont zconfidence}) to be greater than $95\%$, which cuts out 12\% of objects from the preliminary radio sample and 10\% of objects from the preliminary control sample. After redshift selection, $499,418$ radio-quiet galaxies remained in the control sample, $\sim$70\% of the parent MGS, whilst the radio sample contained $17,024$ galaxies. The average error in redshift for both samples is $<0.002$ and the redshift distributions of these two samples were indistinguishable at the 1\% level on the basis of a two-tailed K-S test. 

\subsection{Radio Power}
\label{sec:radioagn}
\citet{Best2005a} suggested that typical radio-loud AGN have powers above $10^{23}$\,W\,Hz$^{-1}$ at 1.4\,GHz and \citet{condon89} shows that P$_{\mathrm{1.4\,GHz}} = 10^{23}$\,W\,Hz$^{-1}$ separates the spiral starburst population from the AGN\,-\,E/S0 population in local galaxy fields.
We therefore restrict the radio sample to galaxies harbouring radio sources with P$_{\mathrm{1.4\,GHz}}> 10^{23}$\,W\,Hz$^{-1}$, which selects 5,119 galaxies (30\%) from the radio sample. Figure \ref{fig:lumredradio} shows that this threshold in radio power is well matched to our redshift selection. 
We name this set of 5,119 galaxies the `radio-loud' sample. 
The mean redshift of the radio sample increased from $\sim0.095$ to $\sim0.136$ when the radio power cut was applied. 

AGN can be split on the basis of large-scale radio structure into FRI type, where radio brightness decreases outwards from the centre and FRII type, with edge-brightened lobes and hot spots \citep{Fanaroff}. FRIIs are more luminous P$_{178\mathrm{\,MHz}}\geq1.3\times10^{26}$\,W\,Hz$^{-1}$ and are rare in a low-redshift sample such as ours (Figure \ref{fig:lumredradio}), since this 178\,MHz power corresponds to P$_{\mathrm{1.4\,GHz}} \approx 3.1\times10^{25}$\,WHz$^{-1}$ for a typical spectral index of $\alpha=0.7$.

\subsection{Removing type 1 AGN}
\label{sec:type1}

In unified AGN models \citep[e.g,][]{Antonucci} the appearance of the central black hole and associated continuum of an AGN differ only in the viewing angle at which it is observed. Sources viewed face on (type 1) show broad emission lines that are absent in those observed edge on and where the broad emission line region is obscured by a dusty torus (type 2). 
Type 1 AGN are excluded from this study, as the optical continuum can be dominated by relativistically boosted non-thermal emission, which may overwhelm measurements of the host galaxy's properties.

The SDSS spectral classification pipeline automatically flags and excludes quasars from the MGS, but we chose to verify its reliability and the relative numbers of type 1 AGN remaining in our sample. We followed the method outlined by \citet{Masci2010} to identify broad line emission. H$\alpha$ or H$\beta$ emission lines exceeding 1000\,km\,s$^{-1}$ (FWHM) with a $S/N > 3$ and H$\alpha$/H$\beta$ EW $>5$\AA\ were classified as broad line. Galaxies with both broad H$\alpha$ and H$\beta$ emission were classified as type 1 AGN. 
3 galaxies of the 5,119 radio-loud sample have broad H$\alpha$ and H$\beta$ emission lines and were removed from the sample. 

184 galaxies (4$\%$) of the remaining 5,116 radio-loud sample have broad H$\alpha$ emission, but do not have broad $H\beta$ emission. 

\citet{Osterbrock1989} classified these objects as type `1.9' AGN, which have substantial but not complete obscuration of the central continuum source. \citet{Kauffmann2003} determine that the contribution to the observed continuum is not significant in these sources, so we retain these within our sample. 
The control sample contained 10 type 1 AGN and 1835 type `1.9' AGN. We removed the type 1 AGN to leave 499,408 galaxies within the control sample. 

\subsection{Removing star-forming galaxies}
\label{sec:sf}
Type 2 AGN have narrow permitted and forbidden lines and their stellar continuum is often similar to normal starforming galaxies. 
\citet{Yun} show a tight correlation between far infra-red luminosity (indicative of star formation) and P$_{\mathrm{1.4\,GHz}}$. This will cause a level of contamination by star-forming galaxies if P$_{\mathrm{1.4GHz}}$ is used as the sole tracer of galaxies hosting a radio-loud AGN. 
We should therefore remove the small subset of radio-loud galaxies in which the radio power arises from star formation (SF) and not from an AGN.

AGN separation from SF galaxies in the local Universe can be achieved via optical emission line ratio diagnostics \citep*[][herein BPT]{BPT}. Emission-line ratios probe the ionizing source: for AGN, non-thermal continuum from the accretion disc around a black hole and in star-forming galaxies (SFGs), photoionization via hot massive stars. 
However, \citet{Best2005b} find no correlation between a galaxy being radio-loud and whether it is optically classified as an AGN. In agreement with this, we found no correlation\footnote{Spearman's rank  correlation result is $\pm$0.06 or less between the radio-flux and optical emission-line flux ratios for our radio galaxy sample.} between radio flux at 1.4\,GHz and the optical line ratios [NII]/H$\alpha$ or [OIII]/H$\beta$ in our sample. 
Hence, a substantial fraction of radio-loud AGN would not be selected using BPT diagnostics, and were we to apply them to the radio sample it would be biased towards radio galaxies with particularly strong optical emission lines.

We instead remove galaxies which are strongly identified as non-AGN, i.e. star-forming.
Despite this method leaving a small fraction of star-forming galaxies which are faint in optical line emission, all radio-loud AGN, whether optically bright or otherwise, will remain in the sample. This decrease in efficiency of the sample is preferential to a drastic decrease in completeness. 
A similar problem was identified by \citet{Sadler2002}, who defined a sample of radio-loud AGN from the 2dFGRS catalogue. They found approximately half of the sample have absorption spectra similar to those of inactive giant ellipticals, and therefore would be mostly missed by  optical AGN emission-line selection.

\begin{figure}
\includegraphics[width=\columnwidth]{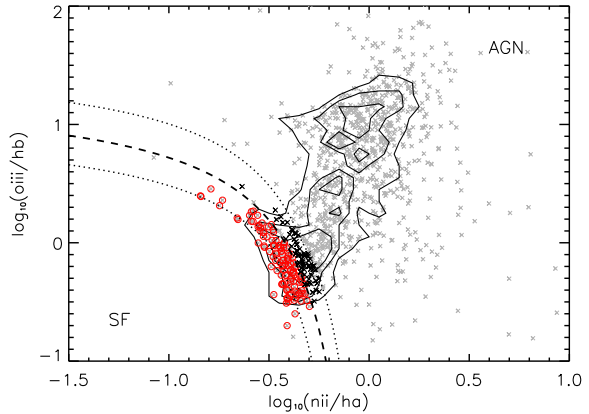}
\caption{BPT diagram for galaxies within the radio-loud sample with emission line ratios. 1,069 galaxies from the 5,116 radio-loud sample have emission line ratios and are plotted as grey crosses where the size of the symbol does not indicate the sizes of the error bars, which vary widely.
The dashed curve (eq. \ref{eq:kauffmann}) indicates the demarcation given by \citet{Kauffmann2003} between optical AGN (above the line) and SF galaxies (below the line).  
766 galaxies plotted have emission line ratios with $>3\sigma$ significance. The density of these galaxies is shown by contours, 85\% of the 766 galaxies lie above the line as optically classified AGN. 
The remaining 15\% (118 galaxies) lie below the curve (red circles) and we classified these as star-forming.
We also show the maximum uncertainty for a point on the Kauffmann separator with S/N of 3 (dotted lines). 
69 objects detected at S/N\,$>3$ (black crosses) lie between the Kauffmann separator and  the upper maximum uncertainty line.
}
\label{fig:bpt_sf}
\end{figure}

In classifying galaxies as AGN or star-forming, we utilized the demarcation criterion of \citet{Kauffmann2003} 

\begin{equation}
\label{eq:kauffmann}
\mathrm{log}(\mathrm{[OIII]}/\mathrm{H}\beta) > 0.61/\{\mathrm{log}([\mathrm{NII}]/\mathrm{H}\alpha) - 0.05\} +1.3 
\end{equation}
plotted as the dashed line on Figure \ref{fig:bpt_sf}.

Figure \ref{fig:bpt_sf} shows the standard line ratio diagnostics for galaxies from the radio-loud sample with all four emission lines catalogued in the SDSS (grey crosses, 1,069 objects of the 5,116). 
766 of these 1,069 galaxies have both optical emission-line ratios at S/N $\geq3$ and their density on the BPT diagram is shown as contours. 118 (11\%) lie below the demarcation line and are marked with red circles (112 out of these 118 have all four emission-lines with S/N $\geq5$). We removed these 118 optically selected SF galaxies from the radio-loud sample, to leave a sample of 4,998 predominantly radio-loud AGN hosts. 

We then estimated the residual contamination expected in the radio AGN sample from star-forming galaxies by plotting the maximum uncertainty for a point on the Kauffmann separator with S/N of 3 (Fig \ref{fig:bpt_sf}, dotted lines). 69 objects detected at S/N\,$>3$ (black crosses) lie between the Kauffmann separator and the upper maximum uncertainty line. Therefore the contamination expected in the radio AGN sample from star-forming galaxies is $\sim11\%$.

Our radio sample contains predominantly FRI galaxies, which are usually hosted by giant elliptical galaxies and on average have weak or no optical nuclear emission lines \citep[and references therein]{Lin}. 
Within our sample of radio-loud AGN, 79\% of objects do not have optical emission line fluxes. We estimate $\approx11\%$ contamination of non-AGN  (e.g. SFG) if galaxies without SDSS emission lines are similar to galaxies with bright line emission. 
5 galaxies without emission line fluxes $>$ 3$\sigma$ lie below the demarcation line in Figure \ref{fig:bpt_sf} and are potentially star forming, but without reliable line information we retain these within our sample.

As discussed by \citet{Best2005b}, a potential shortfall of spectral classification of emission-line radio-loud AGN is that emission-line AGN activity is often accompanied by star formation \citep[e.g.,][]{Kauffmann2003}. This star formation will give rise to radio emission, even if the AGN itself is radio quiet. For these sources, the optical spectrum could still be dominated by a (radio-quiet) AGN leading to classification as an emission-line radio-loud AGN.

\citet{Moric2010} (hereafter M10) derive a population of sources matched from all NVSS galaxies in the Unified Radio Catalog \citep[][]{Kimball}, the SDSS-MGS and IRAS data. The matched sources are divided  into star-forming, composite and AGN using standard BPT diagnostics. Star formation rates are derived via broad-band spectral fitting to the NUV-NIR SDSS photometry, and the \emph{average} fractional star formation/AGN contribution to the radio power is estimated (see their table 2). They find that in 203 composite galaxies, 81.3$\%$ of the total radio power is due to star formation. The variation of this fraction with radio power is not specified. 

Following M10, we defined 206 composite galaxies in our own radio sample (27$\%$ of the emission-line galaxies confirmed at S/N $\geq3$), using the diagnostics of \citet{Kewley} and \citet{Kauffmann2003}. If the average fractional contribution is independent of total radio luminosity, then we estimated an upper limit of $\sim27\%$ of our radio sample may have radio power boosted by star formation but possess a radio-quiet AGN. However, $\lesssim14\%$ of the composite sample defined by  Mori{\'c} have log[P$_{1.4}$(W\,Hz$^{-1}$)] $> 23$. \citet{Mauch2007} find SF galaxies tend to have median log[P$_{1.4}$(W\,Hz$^{-1}$)] = 22.13, whereas AGN have a median log[P$_{1.4}$(W\,Hz$^{-1}$)] = 23.04. Therefore, our luminosity cut will have significantly reduced the numbers of galaxies where star-formation is the principal contributer to the total radio power, and we expect far less than $27\%$ contamination from radio-quiet, optically-loud AGN. 

We also cannot account for the population of `composite' radio-loud host galaxies with ongoing star formation that have been lost from our sample through this selection technique. \citet{Mauch2007} estimate that $\sim 10\%$ of local radio sources may have a starforming spectrum but have radio flux densities dominated by a radio-loud AGN. 

\subsection{Colour bimodality}
\label{subsec:colour}
\begin{figure}
\includegraphics[width=\columnwidth]{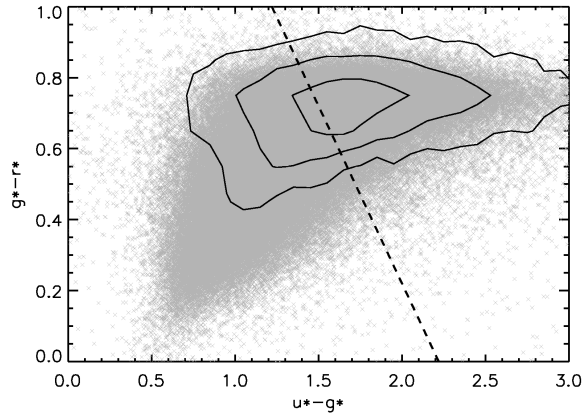}
\caption{Colour-colour plot for the entire control sample (grey) and 34606 visually classified ellipticals from Galaxy Zoo-control sample match (log$_{10}$ contours). The dashed line is $u^{*}-r^{*}\geq2.22$\,mag from \citet{Strateva} and defines a morphological colour separator. 80\% of the Galaxy Zoo-MGS sample lie above the line. The $u^{*}$, $g^{*}$ and $r^{*}$ magnitudes are derived from SDSS columns: {\fontfamily{pcr}\selectfont petrocounts $-$ reddening $-$ kcorr}.}
\label{fig:colourcolour}
\end{figure}

As low-redshift radio-loud AGN are hosted predominantly by elliptical galaxies, we attempted to constrain the control sample to contain only early-type galaxies. The colour distributions of galaxies have been shown to be highly bimodal \citep{Yan2006, Yan2010, Strateva}. Figure \ref{fig:colourcolour} shows a colour-colour plot in $g$*-$r$* against $u$*-$g$* as explored by \citet{Strateva} who define an optimal colour separator\footnote{On a spectroscopically classified galaxy sample, early-types are recovered at $\gtrsim$98\% completeness and $\gtrsim$83\% reliability.} between early and late types of $u^{*}-r^{*}\geq2.22$\,mag. Early-types (E, S0, Sa) populate the upper right region, and late types (Sb, Sc, Irr) occupy the lower left.

In order to test the completeness of this demarcation, we used the 62,190 galaxies visually classified as ellipticals in Galaxy Zoo data \citep{zoo}. We cross-matched these to the control sample, selecting objects within 1\arcsec\ radius.  34,606 matches were found, and 27,569 of these lie above $u$*-$r$*$>$2.22\,mag. Therefore, this demarcation gives $\sim 80\%$ reliability (Fig \ref{fig:colourcolour}, contours), suggesting $\sim 20\%$ of the early-type galaxy population may be excluded through applying this criterion. 
147,275 visually classified `spiral' galaxies from the Galaxy Zoo data are also present in our control sample. 22\% of these lie above $u$*-$r$*$>$2.22, i.e. we expect $\sim22\%$ contamination in the early-type sample.
Of the 499,408 galaxies in the control sample, $\sim40\%$ were classified as early-types, using the \citeauthor{Strateva} colour selection, to leave 199,391 early-type galaxies in the control sample, morphologically selected in colour space at $\sim80\%$ completeness and $\sim78\%$ efficiency based on Galaxy Zoo visual analysis. 

\begin{figure}
\includegraphics[width=\columnwidth]{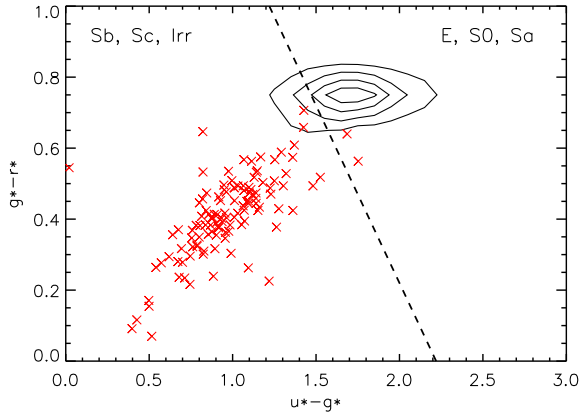}
\caption{Colour-colour plot of radio-loud AGN (contoured, binsize = $0.1\times0.1\,$mag, 4 equally spaced levels, max density contour = 400/bin), and SF galaxies as defined in \S\ref{sec:sf} (crosses) The dashed line is $u^{*}-r^{*}\geq2.22$ from \citet{Strateva} and defines a morphological colour separator.}
\label{fig:radiocc}
\end{figure}

We explore the 4,998 radio-loud AGN selected galaxies in colour space.
Figure \ref{fig:radiocc} shows the radio-loud AGN are predominately early types, with $>70\%$ lying above $u$*-$r$*$>$2.22, i.e. the radio-loud AGN sources are predominantly hosted by ellipticals as defined in colour space. The distribution in colour space is similar to that of the control sample, despite the AGN colours possibly influencing the light. In agreement with this result, \citet{Griffith} found radio-selected AGN have a high incidence of being hosted by early-type galaxies.

The 118 galaxies we flagged as star forming and removed from the radio-loud sample in \S{\ref{sec:sf}} are shown as crosses in Figure \ref{fig:radiocc}, and 98\% of these sit below the line. 
We select the radio-loud AGN above the colour cut, to give 3,516 `early-type' radio-loud AGN hosts.

\subsection{Redshift distributions}
\label{subsec:redshiftdist}
\begin{figure}
\centering
\subfigure[Redshift distributions]{
\includegraphics[scale=0.45]{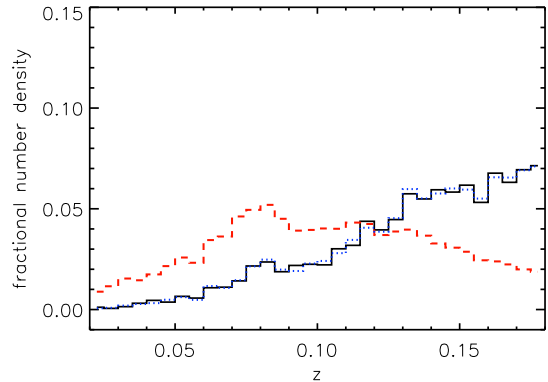}
\label{fig:idl_zscale_lum}
}
\subfigure[Absolute $r^{*}$ magnitude distributions]{
\includegraphics[scale=0.45]{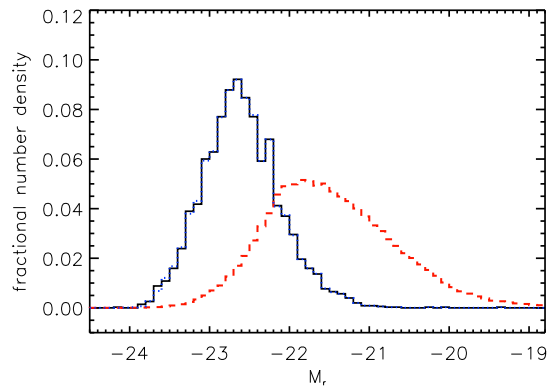}
\label{fig:idl_lum_zscale}
}
\caption{The fractional distributions of (a) redshift and (b) $r^{*}$ band luminosity. The solid black line shows the radio-loud AGN distributions. The red dashed line shows the control sample distribution prior to nearest neighbour matching in redshift-magnitude space. The blue dotted line shows the resultant sample  randomly selected to match the redshift-magnitude distributions of the radio sample.}
\label{fig:idl_scaled}
\end{figure}

At this stage we re-examined the redshift distributions of the samples, using the KS test. The probability of the control sample and the radio-loud AGN being drawn from the same parent sample in redshift was $<0.1\%$: the distributions in redshift differ (Figure \ref{fig:idl_zscale_lum}). 
Evolutionary differences in host galaxy properties should be small over this entire redshift range, however we chose to re-sample the control sample redshift distribution to match that of the radio-loud AGN sample.

We also examined the intrinsic $r^{*}$ band luminosity of the two samples. Figure \ref{fig:idl_lum_zscale} shows the radio sample is globally brighter in $r^{*}$ than the control sample. It has been well established that radio-loud AGN are hosted preferentially in the brightest and most massive elliptical galaxies \citep{Best2005b, Mauch2007}. 
Since we wanted to test whether radio-loud AGN harbour an excess of blue emission in their centres, we chose to re-sample the control sample to have the same $r^{*}$ band optical properties so as to avoid any possible correlation of colour gradient with galaxy luminosity. 

We generated a matched control sample by selecting the 10 galaxies from the whole sample that lie closest to each radio-loud AGN in magnitude-redshift space. The final control sample contains 35,160 galaxies, whose magnitude and redshift distributions are shown as dotted lines in Figure \ref{fig:idl_scaled}. This final selection should allow the distribution of observables in the radio and control samples to be compared directly.

\subsection{Comparison radio sample (CR)}
\label{subsec:crprop}

The criteria applied to the `radio' sample in \S\ref{subsec:redshiftrange}-\ref{subsec:redshiftdist} were also applied to the 5,719 objects in the CR sample (see \S\ref{subsec:crossNVSSFIRST}). 

\begin{figure}
\includegraphics[width=\columnwidth]{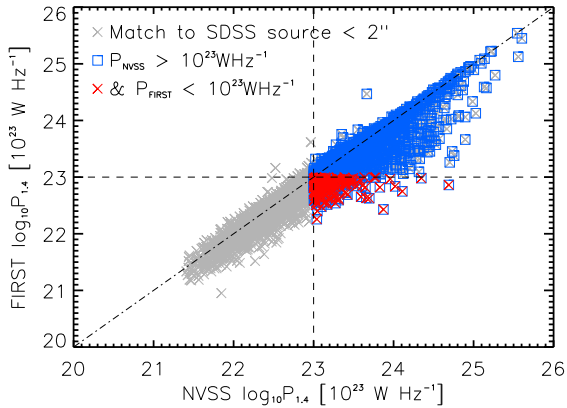}
\caption{FIRST vs NVSS derived luminosities for the 3,737 sources in the candidate CR sample (grey crosses). Those with P$_{\mathrm{1.4\,GHz}}^{\mathrm{NVSS}} > 10^{23}$\,W\,Hz$^{-1}$ are highlighted by blue squares (1,847). Of those, 251 sources with P$_{\mathrm{1.4\,GHz}}^{\mathrm{FIRST}} < 10^{23}$\,W\,Hz$^{-1}$ are shown as red crosses. These are the sources excluded in the radio sample's luminosity cut (\S\ref{sec:radioagn}). }
\label{fig:lumFN}
\end{figure}

We imposed a redshift cut of $0.02<z<0.18$ and restricted the redshift confidence parameter given by SDSS to $>95\%$, leaving 3,727 sources. We examined the CR sample radio powers based on NVSS and FIRST. Figure \ref{fig:lumFN} shows the NVSS and FIRST luminosities for the 3,727 CR sources (grey crosses). Those with P$_{\mathrm{1.4\,GHz}}^{\mathrm{NVSS}} > 10^{23}$\,W\,Hz$^{-1}$ are highlighted by blue squares (50$\%$). We restricted the CR sample to galaxies harbouring radio sources with P$_{\mathrm{1.4\,GHz}}^{\mathrm{NVSS}} > 10^{23}$\,W\,Hz$^{-1}$, which selects 1,847 (50$\%$) of the sample.

Of those 1,847 sources, 251 have FIRST derived powers $< 10^{23}$\,W\,Hz$^{-1}$, and were therefore excluded in the radio sample selection in \S\ref{sec:radioagn}.

We removed 2 type 1 AGN and 48 star-forming galaxies confirmed at S/N $> 3$ from the CR sample, leaving 1,797 radio-loud narrow-line AGN hosts. 4$\%$ of the sample are classified as type `1.9' AGN, with substaintial but not full obscuration of the central source. We applied the Strateva et al. colour cut ($u$*-$r$*$>$2.22\,mag) detailed in \S\ref{subsec:colour} to select 1,295 early-type galaxies from the CR sample.

The control sample was cross-matched with the FIRST-NVSS catalogue to within 2\arcsec and the 5 galaxies found were removed to ensure the comparison control sample is comprised solely of `radio-quiet' AGN elliptical hosts. 199,386 galaxies remain in the CR-control sample. 

We selected the 10 galaxies from the control sample that lie closest to each CR radio-loud AGN in magnitude-redshift space. The final CR-control sample contains 12,950 sources, matched to the CR sample in $r^{*}$-band magnitude and redshift distributions.

\subsection{Summary of Final Samples}
\label{subsec:finalsamples}

\textbf{Radio-loud `early type' AGN (R-AGN) - 3,516 sources} \newline
After matching FIRST to the MGS ($r^{*}_{petro}<17.77$\,mag, $0.02<z<0.18$) within 2\arcsec, we classified as radio-loud AGN hosts with P$_{\mathrm{1.4\,GHz}}> 10^{23}$\,W\,Hz$^{-1}$. Type 1 AGN exhibiting both broad H$\alpha$ and H$\beta$ emission lines were removed from the sample. Optical emission line ratios (if confirmed at 3$\sigma$) were used to remove star-forming galaxies using the  demarcation of \citet{Kauffmann2003}. 
We then created a subset with $u^{*}$-$r^{*}>2.22$\,mag i.e. `early-types' to remain consistent with the colour cut applied to the control sample. 

\noindent\textbf{Control sample - 35,160 sources} \newline 
We removed all galaxies from SDSS-MGS ($r^{*}_{petro}<17.77$\,mag, $0.02<z<0.18$) with FIRST and NVSS  counterparts. Type 1 AGN exhibiting both broad H$\alpha$ and H$\beta$ emission lines were removed from the sample. Galaxies were constrained in colour to bias towards early-type morphology where a comparison with results of the Galaxy Zoo program suggests we have $\gtrsim80\%$ completeness and $78\%$ efficiency. The resultant sample comprises radio-quiet, `normal', predominantly elliptical galaxies. We then selected a subsample to match the redshift and optical $r^{*}$-band magnitude distributions of the R-AGN sample, thus removing any redshift bias and potential correlation of colour gradient with galaxy luminosity.

\noindent\textbf{Comparison radio (CR) - 1,295 sources} \newline
We cross-matched the FIRST-NVSS catalogue derived by \citet{Kimball} with the MGS ($r^{*}_{petro}<17.77$\,mag, $0.02<z<0.18$) to within 2\arcsec, and classified radio-loud AGN hosts as those sources exhibiting  P$_{\mathrm{1.4\,GHz}}^{\mathrm{NVSS}}> 10^{23}$\,W\,Hz$^{-1}$. Type 1 AGN exhibiting both broad H$\alpha$ and H$\beta$ emission lines were removed from the sample. Optical emission line ratios (if confirmed at 3$\sigma$) were used to remove star-forming galaxies using the  demarcation of \citet{Kauffmann2003}. 
We then created a subset with $u^{*}$-$r^{*}>$2.22\,mag i.e. `early-types' to remain consistent with the colour cut applied to the control sample. 

\noindent\textbf{CR-control sample - 12,950 sources} \newline
All selection criteria as above for the control sample, we also removed 5 galaxies within the control sample that were also present in the CR sample. We then selected a subsample to match the redshift and optical $r^{*}$-band magnitude distributions of the CR sample.

We therefore have two radio-loud AGN samples, both with a comparison control sample. The CR sample is derived using both radio catalogues, and is a brighter radio sample (the NVSS flux density completeness limit is $2.5$\,mJy) for comparison with our deeper ($>1$\,mJy), larger but more incomplete R-AGN sample derived solely from the FIRST catalogue.

\section{analysis}
\label{sec:analysis}
The surface brightness profile of an early-type galaxy can be described by the S\'{e}rsic equation 

\begin{equation}
\label{eq:sersic}
I(r) = I(r_{e})e^{-b[(r/r_{e})^{1/n}-1]}
\end{equation}
where $r_{e}$ is the effective radius (scale length) and $I(r_{e})$ is the corresponding effective surface brightness. $b \approx 2n -1/3$ is chosen so $r_{e}$ contains half the light in the galaxy.  
n $\approx$ 4 for bright ellipticals decreasing to n $\approx$ 2 as luminosity decreases \citep{Caon}.
SDSS chooses to fix $n=4$, fitting the `de Vaucouleurs profile', truncating the profile beyond 7\,$r_{e}$.
The model fitted by SDSS has an arbitrary axis ratio and position angle and is convolved with a double-Gaussian representation of the PSF. The fitting then yields, among other properties; the effective radius, $r_{e}$ and error, $\Delta r_{e}$. In general, the error for the $u$ model is greater than other bands. 
It is noted that SDSS's fitting algorithm generates some weak discretization of model parameters, especially in $r_{e}(r)$ and $r_{e}(u)$. Objects with scale lengths lying in discrete bands in $r_{e}$ tend to have poorer goodness of fit. This is not included in the error estimates, $\Delta r_{e}$, which were derived from count statistics on the image. 
In our work with the $r_{e}$ values, we consider $\Delta r_{e}$ rather than {\fontfamily{pcr}\selectfont DeV\_l} (the likelihood of the model fit) due to our reluctance to remove data with poor de Vaucouleurs $u$ band fitting. A poor goodness of fit to the deVaucouleurs profile may be indicative of a cuspy $u$ band and removing such objects could bias results.

SDSS denotes the {\fontfamily{pcr}\selectfont fracDeV} parameter as the fraction of the total flux contributed by the de Vaucouleurs component in a linear combination of a de Vaucouleurs and an  exponential model to find the best fit. 
It is noted that the likelihood values in the $r^{*}$-band are intrinsically poor (all samples have mean values $\sim\,0.01$), but the likelihood ratios still pick out reliable best-fit parameters. 
The fracDeV parameter ($f$ hereafter) is correlated with the S\'{e}rsic index; n\,=\,1 corresponds to $f$\,=\,0, n\,=\,4 corresponds to $f$\,=\,1 \citep[][and references therein]{Kuehn2005}. Following \citet{Vincent2005}, galaxies with $0.5\leq f \leq0.9$ ($2.0 \lesssim$\,n\,$\lesssim 3.3$) are labeled ``de/ex" galaxies and galaxies with $f$\,$\geq 0.9$ (n\,$\gtrsim 3.3$) are labeled ``de" galaxies. 
We chose not to remove objects with $f$\,$<\,0.5$, but instead created subsets of each sample with $f$\,$>\,0.5$ for comparison.

We investigate the colour structure of AGN host galaxies principally through the distribution of $R$, which we define as $R = r_{e}(r)/r_{e}(u)$, which is the ratio of $r$ to $u$ de Vaucouleurs effective radii. This should avoid any intrinsic scale differences in $r$ and $u$ between individual galaxies.

\subsection{Diversity of intrinsic distributions.}

\begin{figure}
\centering
\includegraphics[width=0.45\textwidth]{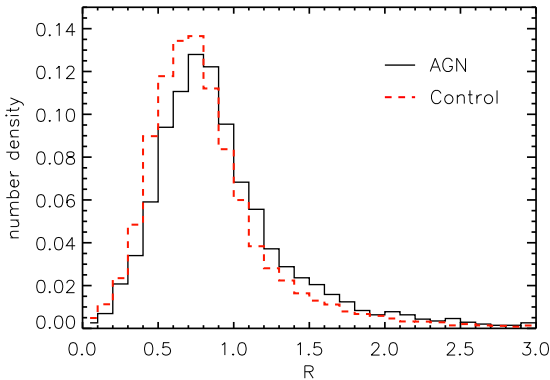}
\caption{Normalised distributions of $R$ for the radio-loud AGN (R-AGN) sample (solid) and the control sample (red dashed).}
\label{fig:dist1}
\end{figure}

The distributions of $R$ for the R-AGN and control sample are shown in Figure \ref{fig:dist1}. Both samples have a few ($<0.001\%$) strong outliers ($R>100$) which significantly disturb the mean. The medians of the R-AGN and control distributions are 0.837 and 0.793 respectively. Using a two-tailed Kolmogorov-Smirnov test we find the probability of the control and the R-AGN sample being drawn from the same distribution to be small ($<0.1\%$).

Similarly, the CR sample's $R$ distribution also shows a higher median compared to the CR-control sample (0.851 and 0.797 respectively). The KS test also finds the probability of the CR and CR-control sample being drawn from the same distribution in $R$ is vanishingly small ($<0.1\%$).

\subsection{Maximum likelihood analysis}
The fractional errors in individual values of $r_{e}(u)$ are relatively large, which makes proving an intrinsic difference between the radio-loud AGN samples and their respective control sample distributions challenging.
We used a maximum likelihood analysis to compare the average values of $R$ for the AGN and control samples, taking into account measurement errors according to \citet{MAC} and \citet{DIANA}, in order to quantify the difference between the distributions of $R$. 

Under the assumption that both the AGN and control samples have normally-distributed intrinsic values of $R$, then the population mean, $\overline{R}$, and  intrinsic dispersion, $\sigma$, of either can be determined from individual values $R_{i}\pm\sigma_{i}$ by minimizing the function

\begin{equation}
\centering{
{\large S} ={\displaystyle \sum\limits_{i}}\left[ \dfrac{(R_{i}-\overline{R})^{2}}{(\sigma^{2}+\sigma_{i}^{2})}+\mathrm{ln}(\sigma^{2}+\sigma_{i}^{2})\right]
}.
\label{eq:smin}
\end{equation}

\begin{figure}
\centering
\subfigure[R-AGN and control]{
\includegraphics[width=0.45\textwidth]{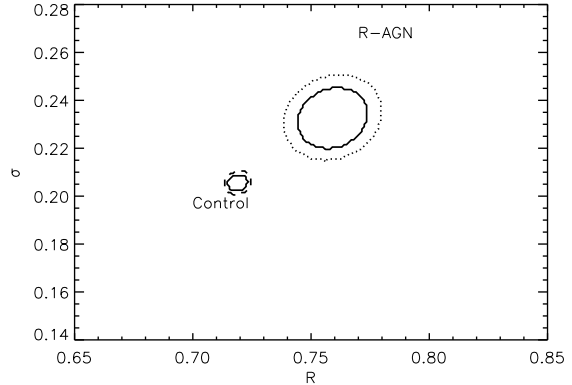}
}
\subfigure[CR and CR-control]{
\includegraphics[width=0.45\textwidth]{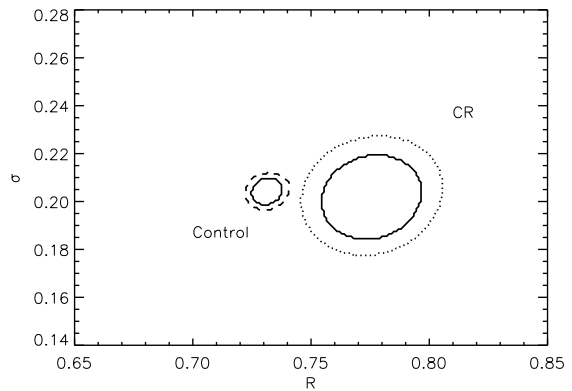}
}
\caption{90\% (solid line) and 99\% (dashed/dotted line) confidence contours of the mean ratio of scale lengths ($R=r_{e}(r)$/$r_{e}(u)$) and intrinsic dispersion for radio-loud AGN galaxies and `normal' early types. The control and radio-loud `early-type' AGN are distinct populations, as are the CR and CR-control samples.}
\label{fig:Rsig}
\end{figure}

Figure \ref{fig:Rsig} shows the results of minimization of Eq \ref{eq:smin} across $\overline{R}$ and $\sigma$ parameter space, for the two samples. The 90\% and 99\% joint-confidence contours for $\overline{R}$ and $\sigma$ are given by S$_{min}+\Delta$S for $\Delta$S$=4.61$ and 9.21 respectively. Table \ref{tab:smin} shows the best fit $\overline{R}, \sigma$ planes for both distributions.
A clear separation between radio-loud `early type' AGN and `normal' control galaxies can be seen for both the R-AGN and CR samples, and we conclude the two are not drawn from the same population at $\gg 99\%$ confidence.

\begin{figure}
\centering
\subfigure[R-AGN and control]{
\includegraphics[width=0.45\textwidth]{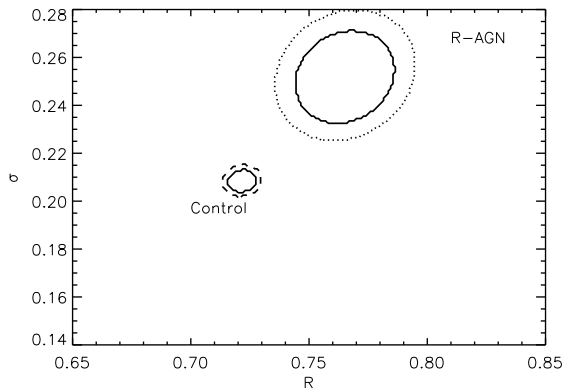}

}
\subfigure[CR and CR-control]{
\includegraphics[width=0.45\textwidth]{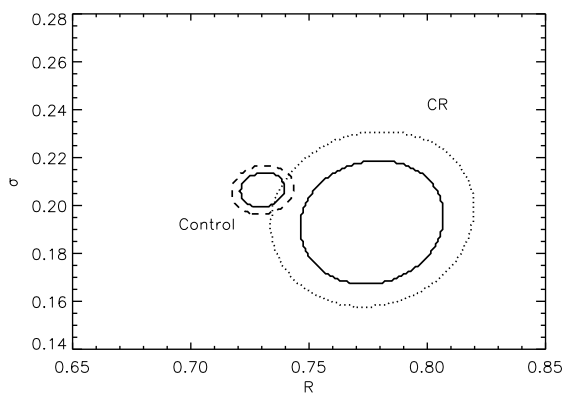}

}
\caption{90\% (solid line) and 99\% (dashed/dotted line) confidence contours of the mean ratio of scale lengths ($R=r_{e}(r)$/$r_{e}(u)$) and intrinsic dispersion for the subset of galaxies with $f>0.5$. The 1,851 radio-loud AGN galaxies and 17,338 `normal' early types are shown in panel (a), the 575 CR galaxies and 6,364 CR-control galaxies are shown in panel (b). The control and radio-loud `early-type' AGN remain distinct populations. The smaller CR and CR-control sample lead to larger confidence contours that overlap, but these remain distinct at $89\%$ confidence.}
\label{fig:Rsigf}
\end{figure}

We applied the cut $f>0.5$ in the $r^{*}$-band, and re-examine the results for both radio-AGN samples. This cut will ensure the reliability of the de Vaucouleur's scale lengths used to derive $R$, and this subset is used as a comparison to the full samples to verify whether galaxies with shapes that differ significantly from the de Vaucouleurs profile perturb our main result.  Table \ref{tab:smin} also shows the fraction of each sample with `good' de Vaucouleurs $r^{*}$-band fits and their $\overline{R}$ values (see Figure \ref{fig:Rsigf}). Although the values do not significantly differ from the full sample, the confidence contours now overlap in the smaller, but more complete, CR sample. However, the R-AGN and control samples are still seen to come from distinct populations.

\begin{table*}
\centering
\begin{tabular}{l ccccc}  
Sample & $\overline{R}$ & $\sigma$ &$f>0.5$& $\overline{R}_{f}$ & $\sigma_{f}$ \\ [0.5ex]    
\hline
\hline
R-AGN   & 0.76$\pm$0.02 & 0.23$\pm$0.01 &53$\%$ & 0.77$\pm$0.02 &0.25$\pm$0.02\\
Control & 0.719$\pm$0.005 & 0.206$\pm$0.003 &49$\%$ & 0.722$\pm$0.006 &0.208$\pm$0.005\\[0.5ex]

CR         & 0.78$\pm$0.02 & 0.20$\pm$0.02 & 44$\%$ &0.77$\pm$0.03 &0.19$\pm$0.03\\
CR-Control & 0.731$\pm$0.007 & 0.204$\pm$0.006 & 49$\%$ &0.730$\pm$0.010 &0.206$\pm$0.007\\
 [0.5ex] 
\hline                         
\end{tabular} 
\caption{Best fit $\overline{R}$ and $\sigma$ corresponding to $90\%$ error, for the R-AGN and CR sample, and their relative control samples. The fraction of each sample with $r^{*}$-band $f>0.5$ is shown, and the resultant mean and standard deviation, $\overline{R}_{f}$ and $\sigma_{f}$. } 
\label{tab:smin} 
\end{table*} 

\section{Discussion}
\label{sec:Discussion}
Our results confirm that radio-loud AGN are hosted by brighter, bigger galaxies on average (see Figure \ref{fig:idl_scaled}). 
Our technique has shown the presence of an AGN is associated with an increase of the scale size of a galaxy in red light relative to blue light. This suggests that radio galaxies have a bluer central bulge and/or more diffusely distributed red light compared to their radio-quiet counterparts. With the analysis so far, we cannot tell whether this difference is from AGN driven or diffuse star-like light within the bulge, or more distributed red light.

 The CR sample is also significantly different in $\overline{R}$ relative to the CR-control sample, so that our R-AGN sample's lower completeness in extended sources does not significantly affect the result. 
The CR sample dispersion is smaller than that of the R-AGN sample, whereas all four control samples have values of $\sigma$ that are similar. We applied a flux cut of $>4$\,mJy to the R-AGN sample (61$\%$), and found $\overline{R}=0.80\pm0.02$ and $\sigma=0.21\pm0.01$. The increased scatter about $R$ is caused by fainter radio sources, which tend to be more heterogenous, and/or have underestimates of errors on $r_{e}$ values in more distant objects.

The subset of CR galaxies constrained to have good $r^{*}$-band de Vaucouleurs fits ($f>0.5$) does not contain enough radio galaxies to show strongly a significant separation from the CR-control sample (see Figure \ref{fig:Rsigf}(b)). We present discussion predominantly on the R-AGN and control sample, which display a clear separation in both Figure \ref{fig:Rsig}(a) and \ref{fig:Rsigf}(a). We concluded the two are not drawn from the same population at $\gg 99\%$ confidence. 
This sample is incomplete to radio sources with lobes $>5.4$\arcsec\ and the population of galaxies with a weak radio AGN core but powerful lobes. However it is large due to the depth of the FIRST survey. The CR results for all Figures in the discussion are similar to the results presented for the R-AGN sample. We find no significant deviation between the two in any subsequent parameter examined, except in $\sigma$, as discussed above.

\begin{figure}
\centering
\includegraphics[width=0.45\textwidth]{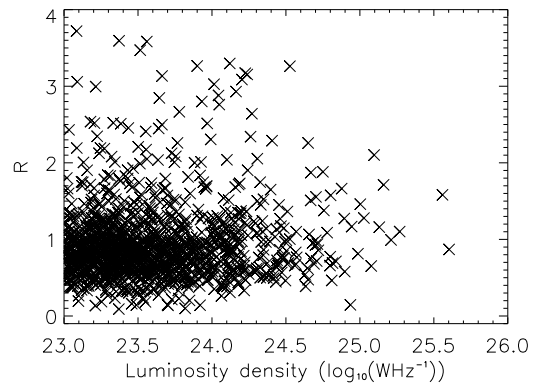}
\caption{Radio luminosity density vs $R$ for the CR sample. We found no correlation ($>99\%$ certainty) between the two.}
\label{fig:lumR}
\end{figure}

\citet[][hereafter MKM99]{Mahab} used a complementary ratio\footnote{$r_{e}(B)/r_{e}(R)$} to demonstrate the ubiquity of inner blue components in a sample of only 30 radio galaxies and 30 control galaxies, and argued that the blue light is due to star formation associated with the presence of a radio source \citep[e.g.][]{anton2009}.
Figure 4 of MKM99 (power at 408\,MHz vs log [$r_{e}(B)/r_{e}(R)$]) shows a trend for more powerful radio galaxies to have steeper colour gradients as indicated by the ratio of scale lengths. 
In contrast to this, when considering only the brightest (P$_{\mathrm{1.4\,GHz}}>10^{25}$\,WHz$^{-1}$) radio sources in the CR sample (for which the powers are more accurate) we find no correlation between radio power and $R$. We also find no correlation over the full range of radio powers and $R$ as shown in Figure \ref{fig:lumR}: the Spearman's rank correlation result is $\pm0.07$ or less between P$_{\mathrm{1.4\,GHz}}$ and $R$. 
We infer that a high value of $R$ (and possible blue central bulge) is not associated with the power of the radio source.

There are several possible scenarios to be considered to explain the larger $R$ in the R-AGN sample. 

\subsection{AGN light}
\label{subsubsec:AGNlight}
Central point-like $u^{*}$ light may be indicative of a blue AGN. 
We investigated the nature of the R-AGN excess $R$, using the goodness of the deVaucouleurs fit. 
The SDSS deVaucouleurs fitting procedure returns the likelihood (between $0$ and $1$) associated with the model from the $\chi^{2}$ fit. If the $u^{*}$-band goodness of fit for the R-AGN sample approaches 1 for low $R$ and 0 for high $R$ then this is indicative of an AGN/core component (as the AGN point-like component would perturb a deVaucouleurs fit) driving a blue excess and high $R$. 
If no structure is seen in $R$ vs deVaucouleurs $u^{*}$-band goodness of fit in either sample, then any blue excess is more likely driven by some sort of diffuse starlight in the central bulge. 

The $u^{*}$-band data for both the R-AGN and control samples are generally well fit by a deVaucouleurs profile (the R-AGN/control median likelihood values are 0.86 and 0.83 respectively), and there is no trend for the $u^{*}$-band or $r^{*}$-band fits to be worse at large $R$ in the R-AGN than the control sample. We can conclude that the high $R$ in the R-AGN sample is not due to any point-like AGN component.

\subsection{Radio AGN triggering star formation}

It is generally agreed that AGN feedback regulates the supply of cold gas for star formation by supplying energy to the ISM and IGM. Some radio-loud star-forming AGN show a connection between the suppression of star formation and the strength of the radio jets, by heating and expelling the surrounding gas \citep{Nesvadba}. \citet{Schawinski09} looked at a sample of low redshift SDSS early-type galaxies for which late-time star formation is being quenched. They found that molecular gas disappears less than 100\,Myr after the onset of accretion onto the central black hole. These galaxies were not associated with radio jets, but show that low-luminosity AGN episodes are sufficient to suppress residual star-formation in early-type galaxies. 

\begin{figure}
\centering
\includegraphics[width=0.45\textwidth]{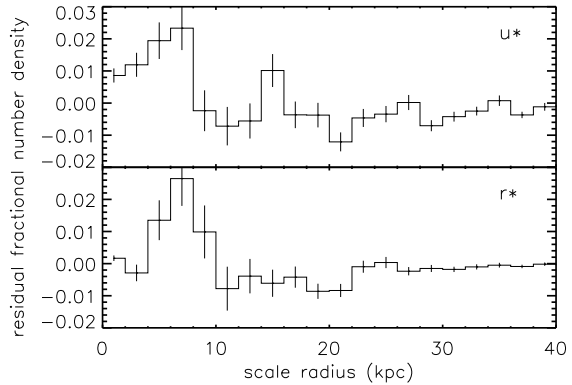}
\caption{Residual fractional number density distributions (R-AGN - control), in 2\,kpc bins, for the $u^{*}$-band and $r^{*}$-band scale radii. There is a higher fraction of blue cores in the R-AGN sample as compared to the control sample at $r_{e}\sim10$\,kpc. This plot does not demonstrate the correlation between the red and blue scale radii (see Figure \ref{fig:rdev}). The measurement errors are larger in the $u^{*}$ band which may account for the higher variability in the residuals.} 
\label{fig:residual}
\end{figure}

\begin{figure}
\centering
\includegraphics[width=0.45\textwidth]{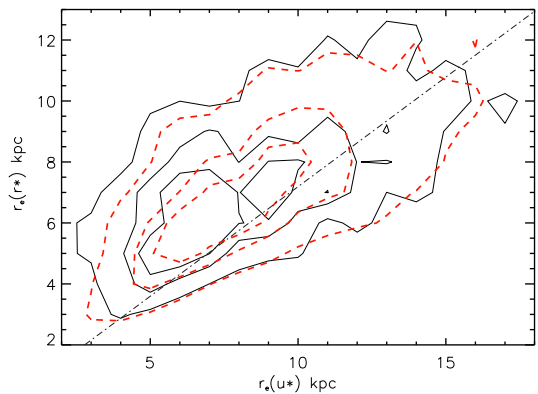}
\caption{Normalised number densities of $r_{e}(u^{*})$ and $r_{e}(r^{*})$ for the R-AGN (solid) and control sample (red dashed). The dotted\,-\,dashed line is the locus $r_{e}(r^{*})=0.719 r_{e}(u^{*})$, galaxies above this line become bluer inward. Contours show the fractional number densities of each sample with levels at 0.004, 0.008 and 0.011. There is a larger fraction of the R-AGN sample above the line, possibly denoting a tendency to become bluer inward more often than the control sample.}
\label{fig:rdev}
\end{figure}

The residual fractional number density distribution of scale lengths are shown in Figure \ref{fig:residual}, for the control sample subtracted from the R-AGN sample. There is a clear excess of blue cores ($0.1-10$\,kpc) in the R-AGN sample as compared to the control sample. There are also more small $r_{e}(r^{*})$ in the R-AGN sample, ranging from $4-10$\,kpc. This plot does not demonstrate the correlation between the red and blue emission, which is shown in Figure \ref{fig:rdev}, but it does demonstrate that the presence of a radio AGN seems not to have suppressed star-formation in the central regions of its host galaxy.

Figure \ref{fig:rdev} plots $r_{e}(u^{*})$ against $r_{e}(r^{*})$ for the R-AGN sample (black) and the control sample (red dashed). The dotted\,-\,dashed line denotes the locus $r_{e}(r^{*}) = 0.719 r_{e}(u^{*})$ as defined by the control sample (see Table \ref{tab:smin}).

At $r_{e}(r^{*})>0.719 r_{e}(u^{*})$, the surface brightness in $u^{*}$ increases more rapidly toward the centre of a galaxy than in $r^{*}$ as compared to the mean of the control sample, i.e. half the blue light from the galaxy is contained in a smaller region than half the red light, implying the galaxy becomes bluer inward. 

The distribution of number densities in Figure \ref{fig:rdev} shows radio galaxies appear to become bluer inwards more often than the control galaxies (66\% of the R-AGN sample are above the locus $r_{e}(r) = 0.719\,r_{e}(u)$, compared to 60\% of the control sample).

There are fewer blue smaller cores in the control sample as compared with the R-AGN sample (see also Figure \ref{fig:residual}), implying there may be a blue excess near the AGN core for some of the radio galaxies. However, the higher $R$ values of the R-AGN sample also seems to be driven by a higher scale length in $r^{*}$, seen in the difference between the inner two contour levels of each sample (see also \S\ref{subsubsec:redenv}), so we cannot definitively conclude that the increased $R$ in the R-AGN sample is due to star formation within the central few kpc. Star-formation within the central few kpc would contradict feedback models which predict the suppression of star formation near an AGN.

\subsection{Star formation in the bulge}  
\citet{Shabala} found that radio sources in massive hosts are re-triggered more frequently than their less massive counterparts, suggesting that the onset of an AGN quiescent phase is due to fuel depletion. AGN activity is therefore promoted by an increase in gas in the centres of the galaxies, which may imply a link between the AGN radio phase and star formation in the bulge ($<10$kpc) of the host galaxy. 

It has been suggested that AGN activity and a major episode of star formation in radio-loud galaxies is triggered by the accretion of gas during major mergers and/or tidal interactions. However AGN activity is initiated later in the merger event than the starburst \citep[e.g.][]{Schawinski07, emonts, tadhunter}. Our spectroscopic selection rules out major merger events (see \S \ref{subsubsec:mgs}), but residual star formation may still be present on global scales. 

Observationally, there is a strong link between AGN and starbursts \citep[][and references therein]{Shin}. \citet{Kauffmann2003} found powerful optical AGN (as classified by the strength of the [O\text{\small III}] emission $L_{\mathrm{[OIII]}}>10^7 L_{\odot}$) predominantly reside in `young bulges'. Recent star formation can provide up to $25-40\%$ of the optical/UV continuum in radio galaxies at low and intermediate z e.g. \citep[e.g.,][]{Holt, tadhunter}. 

We found a higher fraction of R-AGN galaxies have $r_{e}(u^{*})<10$\,kpc as compared to the control sample (see Figure \ref{fig:residual}), perhaps indicative of star formation in the outskirts of the bulge, fuelled by gas expelled from the central regions by the AGN. 

\subsection{Diffuse red emission}
\label{subsubsec:redenv}
An alternative interpretation of a larger $R$ is that the R-AGN sample has more distributed red light. Radio-loud `early-type' galaxies predominantly reside in elliptical galaxies, which are well known to be redder than late-types \citep[e.g.,][]{Strateva}. Diffuse red emission would cause a larger deVaucouleurs scale radius in the $r^{*}$ band and an increased $R$.

Figure \ref{fig:rdev} suggests the increased $R$ in the R-AGN sample may also be attributed to more diffuse red light; between $r_{e}(u^{*}) \sim4-10$\,kpc, $r_{e}(r^{*})$ is on average higher for the radio population. This is also seen in the lower panel of Figure \ref{fig:residual}.

Our results show a difference in the distributions of red and blue light in R-AGN and normal `early-type' galaxy populations, though a high value of $R$ is not a property of every active galaxy. We found the higher $R$ in the R-AGN sample is contributed to by a higher fraction of radio galaxies harbouring small blue cores but also an increase in the numbers of radio galaxies with more diffuse red light as compared to the control sample. 
Given that star formation proceeds over a longer timescale than radio activity, this disfavours the idea that all galaxies undergo short bursts of radio activity, but rather implies that a subset have the predisposition to become radio-loud.

\section{Conclusion}
\label{conclusion}

We cross-matched low-redshift ($0.02<z<0.18$) data from the SDSS MGS ($r^{*}_{petro}<17.77$\,mag) and FIRST to within 2\arcsec, deriving a radio sample of galaxies at $>99\%$ efficiency and $>72\%$ completeness. Radio luminosities were in the range $10^{23}-10^{25}$\,WHz$^{-1}$. Type 1 AGN were removed from the sample (identified via $H\alpha$ and $H\beta$ broad line characteristics), 4$\%$ of the sample are classified as type `1.9' AGN, with substantial but not full obscuration of the central source. Contamination from SFGs (identified via optical emission line ratios) in the radio sample is expected at $\sim11\%$. 
A control sample was defined from SDSS sources with no match to a FIRST source within 2\arcsec\ of their optical core, providing a sample of $>99\%$ efficiency and $>92\%$ completeness. 
At 80\% reliability, the demarcation $u^{*}-r^{*}>2.22$\,mag selected `early-type' galaxies in both samples. Samples were matched in $r^{*}$-band magnitude and redshift distributions and final sample sizes were 3,516 radio-loud AGN galaxies (R-AGN) and 35,160 control galaxies. 

We also created a complementary flux-limited sample through cross-matching with NVSS (the CR sample). The same cuts were applied to derive a radio-loud `early-type' AGN sample, except we used the more all-encompassing NVSS flux estimates to cut in radio luminosity, thereby retaining the populations of galaxies with weak/no core radio emission but bright, extended radio lobes. This sample had higher completeness for comparison with the R-AGN sample. A control sample was defined from the SDSS sources with no match to a NVSS-FIRST source within 2\arcsec\ of their optical cores. We further considered only sources where fracDeV\,(f) $>0.5$ to restrict all four samples to having good $r^{*}$-band deVaucouleurs fits. 

We investigated the colour structure of AGN host galaxies through $R$, the ratio of $r^{*}$ to $u^{*}$ de Vaucouleurs effective radii and used maximum likelihood analysis to quantify the degree of difference in the distribution of $R$ between samples. We concluded the radio (R-AGN/CR) samples are not drawn from the same population as their radio-faint control samples at $\gg99\%$ confidence: the presence of an AGN increases the scale size of a galaxy in red light relative to blue light, on average.

Our result does not appear to be driven by the presence of blue AGN in the radio-loud samples since the goodness of the de Vaucouleurs fits does not become worse as $R$ increases. We found no structure in $R$ vs $u^{*}$-band goodness of fit in either radio sample. 
We found an excess of blue cores in radio-loud galaxies as compared to radio-quiet, `early-type' galaxies, implying the increased $R$ may be due to star formation in the central few kpc, in contrast with feedback models which predict the suppression of star formation near an AGN. Spectroscopic selection of our samples rules out major merger events, as starbursts can be triggered by the accretion of gas/tidal interactions. 
We also found radio AGN hosts to have larger red scale lengths in relation to their blue light and note this to be a contributing factor in an increased $R$. We cannot definitively discern whether a small blue core or larger distribution of red light is the driving factor in this result. 

Given the longer timescale for star formation than radio activity, our results imply a subset of galaxies have the predisposition to become radio-loud, rather than all galaxies undergoing bursts of radio activity at some stage in their lifetimes. 

\section*{Acknowledgements} 
EM gratefully acknowledges support from the UK Science and Technology Facilities Council and thanks Luke Davies and James Price for helpful comments. We thank the anonymous referee for helpful and insightful comments that have resulted in an improved analysis. In undertaking this research, we made extensive use of the {\sc topcat} software \citep{Taylor}. 
Funding for the SDSS has been provided by the Alfred P. Sloan Foundation, the Participating Institutions, the National Science Foundation, the U.S. Department of Energy, the National Aeronautics and Space Administration, the Japanese Monbukagakusho, the Max Planck Society, and the Higher Education Funding Council for England. This research makes use of the NVSS and FIRST radio surveys, carried out using the National Radio Astronomy Observatory Very Large Array: NRAO is operated by Associated Universities Inc., under cooperative agreement with the National Science Foundation.


\end{document}